\documentclass[12pt]{article}
\usepackage{epsfig}
\title{A Phenomenological Estimate of the Binding Energy of Heavy Dimesons
      \thanks{submitted to {\it Few Body Systems}}}
\author{D. Janc\thanks{E-mail: damjan.janc@ijs.si}\ \ and
        M. Rosina\thanks{E-mail: mitja.rosina@ijs.si} \\[12pt]
{\sl Faculty of Mathematics and Physics, University of Ljubljana,} \\
{\sl              Jadranska 19, P.O.~Box 2964, 1001 Ljubljana, Slovenia,}\\
{\sl and J.~Stefan Institute, Ljubljana, Slovenia}}
\date{}

\begin{document}
\def\MeV{\,{\rm MeV}}
\def\GeV{\,{\rm GeV}}
\def\fm{{\,\rm fm}}
\def\half{{\textstyle\frac{1}{2}}}
\def\Frac#1#2{{\textstyle\frac{#1}{#2}}}
\def\sqrthalf{{\textstyle\frac{1}{\sqrt{2}}}}

\maketitle

\begin{abstract}

A  phenomenological estimate is derived such that
the binding energies of dimesons are expressed as combinations
of masses of different mesons and baryons. The estimate is almost
model-independent, the only major assumptions being
that the wave functions of the two light quarks in $\Lambda_c$,
$\Lambda_b$ and in the $\bar{c}\bar{c}qq$ and $\bar{b}\bar{b}qq$
dimesons are very similar, and that for heavy quarks the $QQ$ interaction
is half as strong as the $Q\bar{Q}$ interaction. 
We get $\bar{b}\bar{b}qq$ (I=0, J=1)
bound by about 100 MeV and $\bar{c}\bar{c}qq$  unbound.

\end{abstract}

\section{Introduction}

The constituent quark model has been rather successful in describing 
the properties of individual hadrons \cite{SB1,SB2,Graz}. The 
extrapolation
to two-hadron systems is, however, still rather uncertain and the
predictions of various quark models sometimes differ dramatically. 
One source of discrepancy is our limited knowledge about the ``correct''
effective quark-quark interaction, and another source are different
choices of model parameters designed to fit different subsets of hadrons.
Even the calculation of the nucleon-nucleon interaction in the model 
with two three-quark clusters is quite unreliable.

Much can be learned by studying a simpler two-hadron system:  
two heavy mesons.
The lowest states should be very narrow (longlived)
since they can decay only weakly.
Although their width is only  $\sim 10^{-3}$ eV (corresponding to a
lifetime of $\sim$1 ps) they are difficult to detect
because of a low production cross section.
Nevertheless, they are interesting theoreticaly, to confront
different models. The detailed calculations in the literature
\cite{SB3,BS1} rely on particular quark models, therefore we attempt
an almost model-independent phenomenological estimate.

In Sect.2 we present our phenomenological estimate of the binding 
energy
in order to support the expectation that some heavy dimesons may be 
bound.
The estimate is mainly based on the assumption
that the wave functions of the two light quarks around the heavy quark
in $\Lambda_c$, $\Lambda_b$ 
and around the antidiquark in the 
$\bar{c}\bar{c}qq$ and $\bar{b}\bar{b}qq$
dimesons are very similar. This assumpton implies that the heavy 
antidiquark in a colour triplet state acts just like a very 
heavy quark 
and that the $1/m$ corrections are neglected. 

This assumption is supported by the experimental observation
that the masses of many meson or baryon pairs with $c$ quark replaced 
by $b$ quark
differ by the same amount (essentially by the quark mass difference):
$\tilde{B}-\tilde{D}=3341\MeV,\,\tilde{B}_s-\tilde{D}_s=3328\MeV,\,
\Lambda_b-\Lambda_c=3340\MeV.$ 
\footnote{In this paper we shall denote the masses
of particles just by their names, and the tilde denotes
a hyperfine average 
(e.g. $\tilde{D}=\frac{1}{4} D + \frac{3}{4} D^*$).}

The second assumption is that for heavy quarks the $QQ$ interaction
is half as strong as the $Q\bar{Q}$ interaction. This is true for the
one-gluon-exchange interacton and popular for the confining interaction.
In potential models with 2-body interaction it is difficult to avoid
the $\lambda\cdot\lambda$ factor in the interaction without
closing all open channels for white clusters. One should note also that
for heavy quarks the one-pion-exchange potential does not apply.

In Sect.3 we show that several refinements may give corrections 
of $\pm20$ MeV but they do not change the qualitative features
and conclusions.

We comment the results in the Conclusion where we explain why  
some previous model calculations are in agreement with
our phenomenological results and why some other calculations 
which disagree are irrelevant.

\section{The phenomenological relation for the binding energy of 
dimesons}

The results in this section, Eqs. (\ref{binding}) and (\ref{result}), 
are the main message of our paper.
In next section we only scrutinize the assumptions and neglected
terms and estimate the minor corrections. While in the present section
the calculations are very simple, essentially interpolations between
experimental hadron masses, the corrections in next section rely 
strongly
on detailed constituent quark model calculations \cite{Janc}.

We derive  phenomenological relations between dimesons, mesons and 
baryons by making two assumptions, (i) the equality of the wavefunction
of two light quarks in the dimeson and in a heavy baryon, and
(ii) the similarity between the subsystem of two heavy quarks 
in the dimeson and the quark-antiquark system in a heavy meson.

The first assumption implies that the heavy antidiquark in a colour
triplet state acts just like a very heavy quark and that the 
$1/m$ corrections are neglected; however, there is no prejudice
for the light-light subsystem in favour of the 
one-gluon-exchange or one-Goldstone-boson-exchange interaction.

The second assumption implies that effective interaction between
two heavy quarks in colour triplet state is half as strong as the
effective quark-antiquark interaction in colour singlet state;
therefore the binding energies of diquarks are weaker than those of
mesons, and the distances increase.
We do not prejudice the radial dependence of the effective interaction
(the flavour-dependent spin-spin interaction is treated later 
as a perturbation). 

We call the $u$ and $d$ quarks $q$ and the dimesons 
$(\bar{b}\bar{b}qq)=T_{bb},\,(\bar{c}\bar{c}qq)=T_{cc}$
(the symbol $T$ reminds of the terminology {\em tetraquarks}).

The binding energy $E_{b\bar{b}}$ of a quark and antiquark in a meson is a 
function
of the reduced mass only, e.g. $\Upsilon=b+b+E_{b\bar{b}},\,\,
E_{b\bar{b}}=F(m=b/2)$. For the diquark $bb$ the Schr\"odinger equation
is similar as for the $b\bar{b}$ meson with twice weaker interaction
\begin{equation}
\left[\frac{p^2}{2(b/2)} + V_{bb}\right]\psi =
\half\left[\frac{p^2}{2(b/4)} + V_{b\bar{b}}\right]\psi = E_{bb}\psi
\end{equation}
To get the similarity, we have mimicked the kinetic energy with half
smaller reduced mass. The binding energy is then $E_{bb}=\half F(b/4)$.

Now we compare the following hadrons
\begin{eqnarray}
T_{bb}    &=& 2b + 2q + E_{bb} + E_{qqQ} \nonumber\\
\Upsilon  &=& 2b + E_{b\bar{b}} \nonumber\\
\Lambda_b &=& b + 2q + E_{qqQ},
\end{eqnarray}
where $E_{qqQ}\approx E_{qq(\bar{b}\bar{b})} \approx E_{qqb}$ is the
potential plus kinetic energy contribution of the two light quarks 
in the field of a "heavy quark". The quantity $E_{qqQ}$ is taken from
the experimental data on $\Lambda_c$ and $\Lambda_b$ and we do not need
to calculate it.

We obtain the phenomenological relation
\begin{equation}
T_{bb}=\Lambda_b+\half\Upsilon+\delta E_{bb},\qquad
       \delta E_{bb}=\half[F(b/4)-F(b/2)].
\end{equation}

An analogous comparison gives for the charm-charm dimeson
\begin{equation}
T_{cc}=\Lambda_c+\half J/\psi+\delta E_{cc},\qquad
       \delta E_{cc}=\half[F(c/4)-F(c/2)].
\end{equation}

The binding of the $(I=0,J=1)$ dimesons is expressed with respect
to the corresponding thresholds. In order to be unambiguous
we give in Appendix all experimental data and all references to model
parameters which we need now or later.
\begin{eqnarray}
\triangle T_{bb}&=&\Lambda_b+\half\Upsilon-B-B^*+\delta E_{bb}
                  =-250\MeV+\delta E_{bb}, \nonumber\\
\triangle T_{cc}&=&\Lambda_c+\half J/\psi-D-D^*+\delta E_{cc}
                  =-42\MeV +\delta E_{cc}
\label{binding}
\end{eqnarray}

\begin{figure}[htb]
\centering
\epsfig{file=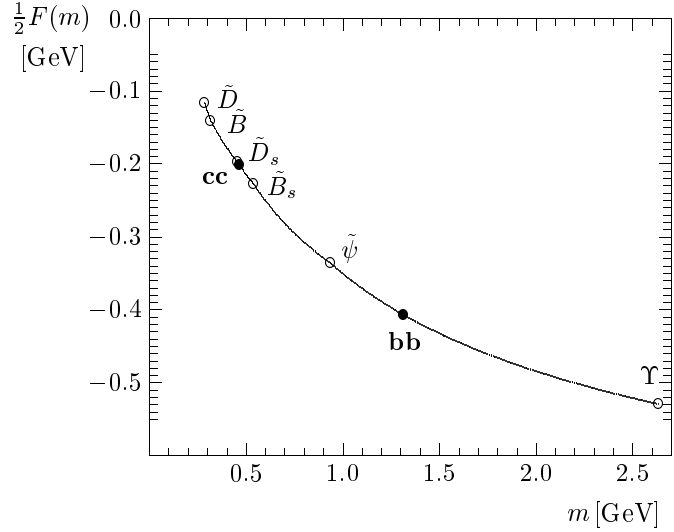,width=250pt}
\caption{%
Interpolation for the $bb$ binding energy -- $BD$ parameters
(The curve for AL1 parameters is similar.)}
\label{Inter}
\end{figure}

Now comes an important idea how to obtain phenomenologically the
``corrections'' $\delta E$. In Fig.(\ref{Inter}) we
plot the phenomenological binding energies obtained from
experimental meson masses and from a popular sets of quark masses,
as a function of the reduced masses of the $b\bar{b},c\bar{c},
b\bar{s},c\bar{s},b\bar{q}$ mesons. We chose the quark masses
proposed by Bhaduri et al.($BD$) \cite{BD} and by Silvestre-Brac 
($AL1$) \cite{SB2}, collected in Appendix; because of similarity
between the two sets only the Fig.(\ref{Inter}) for the set ($BD$)
is presented. The results are only weakly dependent on quark masses.
We then interpolate in order to obtain the values
of $F$ for the scaled reduced masses of diquarks. 
The idea to interpolate between experimental meson masses has
been used before \cite{LRP}, but their construction of dimesons is 
quite different: a light diquark and a heavy antidiquark bound as a meson;
however, the assumption of a light diquark is less justified than
our construction with full freedom for the two light quarks.

For the interpolation we use the following data
\begin{eqnarray}
\half E_{b\bar{b}} &=& \half\Upsilon-b\equiv \half F(\half b)
=-529 \MeV\;(BD),\qquad -497 \MeV\;(AL1), \nonumber\\
\half E_{c\bar{c}} &=& \half \tilde{\psi}-c\equiv \half F(\half c)
=-336 \MeV\;(BD),\qquad -302 \MeV\;(AL1),\nonumber\\
\half E_{c\bar{s}} &=& \half(\tilde{D}_s-c-s) \nonumber\\
&\equiv& \half F((c^{-1}+s^{-1})^{-1}) 
=-197 \MeV\;(BD),\qquad -168 \MeV\;(AL1). 
\end{eqnarray}
Here we used the hyperfine averages 
$\tilde{\psi}=\frac{3}{4}J/\psi+\frac{1}{4}\eta_c$ and similarly for 
$D_s$;
for $\Upsilon$ we take the $S=1$ state since $\eta_b$ is unknown and 
the
splitting is anyway small.

For the interpolation between $\Upsilon$ and $\tilde{\psi}$ 
we plot in Fig.(\ref{Inter}) curves of the form 
$\sigma m^{-1/3}+\tau$. Such interpolation corresponds to the scaling
for the linear quark-antiquark potential; it also gives a rather
straight line for phenomenological binding energies if plotted
against the abscissa $m^{-1/3}$. We get 
\begin{eqnarray}
BD:  \qquad \half F(\Frac{1}{4}b)&=&\half F(1315\MeV)=-407 \MeV. \nonumber\\
AL1: \qquad \half F(\Frac{1}{4}b)&=&\half F(1307\MeV)=-375 \MeV.
\end{eqnarray}
There is some uncertainty in the interpolation. 
The extreme choices of linear interpolation of $F$ versus $1/m$ 
(or $m$)
give 15 MeV stronger (27 MeV weaker) binding, respectively,
suggesting an error $\pm15 \MeV$.
The binding energy $\triangle T_{bb}$ is  then
\begin{eqnarray}
BD: \qquad \delta E_{bb}&=&+122\pm15 \MeV,\qquad
          \triangle T_{bb}=-128\pm15 \MeV \nonumber\\
AL1:\qquad \delta E_{bb}&=&+122\pm15 \MeV,\qquad
          \triangle T_{bb}=-128\pm15 \MeV 
\label{result}
\end{eqnarray}

These values are very close to the result $\triangle T_{bb}=-131\MeV$
of a detaild 4-body calculation with $BD$ interaction \cite{SB3}.

For $T_{cc}$ it is even easier to interpolate since the reduced mass 
$c/4$ happens to lie very close to the reduced mass of the 
$\tilde{D}_s=c\bar{s}$ meson so that 
\begin{equation}
F(\Frac{1}{4}c)=F(467\MeV)\approx F((c^{-1}+s^{-1})^{-1})=F(454\MeV)
\end{equation}
for $BD$ and similar for $AL1$.
However, the choice of the strange quark mass
brings some slight additional model-dependence. 

We get
\begin{eqnarray}
BD: \qquad \delta E_{cc}&=&+139 \MeV,\qquad
          \triangle T_{cc}=+97 \MeV \nonumber\\
AL1:\qquad \delta E_{bb}&=&+134 \MeV,\qquad
          \triangle T_{bb}=+92 \MeV 
\end{eqnarray}
Therefore we expect rather reliably that the $DD^*$ system $T_{cc}$
is unbound.

It is amusing to perform an analytic calculation of $\delta E$
by mimicking the $\alpha/r+\kappa+\lambda r$ potential by a 
logarythmic 
potential $U \ln(r/r_0)$. The ground state energy of mesons is 
\begin{equation}
E=F(m)=U\left(\epsilon-\half\ln[2mr_0^2U/\hbar^2]\right)
\end{equation}
where $m$ is the reduced mass and $\epsilon=1.0443$ is obained by the 
numerical solution of the Schr\"odinger equation in dimensionless form.
Then we obtain for any quark $Q=c,b,...$
\begin{equation}
\delta E_{QQ}=\half[F(Q/4)-F(Q/2)]= \Frac{\ln 2}{4}\,U = 127\MeV
\end{equation}
which is surprisingly close to the above phenomenological estimates
for the $bb$ and $cc$ diquarks. The strength $U=733\MeV$ is taken from
ref. \cite{Log}.

\section{Refinements}

In this section we make several refinements and corrections in order to
test our assumptions and approximations. It turns out that the 
refinements 
may change the results only by 10-30 MeV so that our qualitative 
conclusions
remain valid.

\subsection{Sensitivity to quark masses}

We repeat the calculation with a rather different choice of masses.
We explore smaller masses. We propose the following phenomenological
choice in which we assume that the wave functions of both partners 
are the same.
Then mass differences of mesons correspond to mass difference of 
quarks,
and the ratio of hyperfine splits corresponds to the mass ratio 
of quarks.
\begin{eqnarray}
b-c\approx\tilde{B}-\tilde{D}=3341\MeV,\qquad b &=& 4941\MeV 
\nonumber\\
b/c\approx\frac{D^*-D}{B^*-B}=3.087,\qquad c &=& 1600\MeV  \nonumber
\end{eqnarray}
The results (using the linear interpolation of $E$ versus $m^{-1/3}$) 
are 
$$
E_{bb} = -128 \MeV,\qquad \delta E_{bb}=83 \MeV,\qquad
\triangle T_{bb}= -167 \MeV
$$
As we see, smaller quark masses
yield a somewhat stronger binding. The reason is indirect, the 
separation
of meson masses in quark masses plus binding energy is to some extent
arbitrary, smaller quark masses mean less negative binding energy 
and therefore a smaller $\delta E$.

For the strange quark such suggestions are very uncertain since the
assumption of equal wave functions is not justified. Nevertheless, 
we take
the suggestions as a guidance (we take for the light quark 
$q=320\MeV$.)
\begin{eqnarray}
c-s\sim\tilde{D}-\tilde{K}=1179\MeV,\qquad s &\sim& 421\MeV \nonumber\\
s/c\sim\frac{D^*-D}{K^*-K}=0.355,\qquad s &\sim& 568\MeV    \nonumber\\
s-q\sim\tilde{D}_s-\tilde{D}=103\MeV,\qquad s &\sim& 420 MeV\nonumber\\
s-q\sim\tilde{B}_s-\tilde{B}= 90\MeV,\qquad s &\sim& 410 MeV \nonumber
\end{eqnarray}
and choose for the phenomenological calculation $s= 420 \MeV$.
Since the reduced mass for the $cc$ diquark $m=c/4=400\MeV$
is close to the reduced masses for $c\bar{s}$ (333 MeV) 
and $b\bar{s}$ (387 MeV) we interpolate linearly and we obtain
$$
E_{cc} = +19 \MeV,\qquad \delta E_{cc}=85 \MeV,\qquad
\triangle T_{cc}=+43 \MeV
$$
The $DD^*$ dimeson is unbound also for these smaller masses.
We conclude that there is some model-dependence on masses, but not
too strong.

\subsection{The spin-spin interaction}

Since the $bb$ diquark is colour antisymmetric and orbital symetric
it must be spin symmetric (S=1) and is analogous to $\Upsilon$.
On the other hand, to get $F(b/4)$ we had to interpolate between
$F(b/2)$ and $F(c/2)$. Since the spin-spin interaction is flavour
dependent it scales differently with masses as the spin-independent
part of interaction and we have interpolated between hyperfine 
averages.
Therefore $F_{S=1}(c/2)$ corresponding to the $bb$ system should be 
lifted by 
$\half\cdot\frac{1}{4}(J/\psi-\eta_c)\,(c/b)=5\MeV$ (for $BD$)
and $F(b/4)$ even less, which is negligible. (Since $\eta_b$ is still
unknown, we have actually used $\Upsilon$ instead of $\tilde{\Upsilon}$
and the lifting is still smaller).

For the $cc$ diquark the shift is sligtly larger. In the hyperfine 
average,
$E_{cc}$ is close to $\half E_{c\bar{s}}$. The splitting 
$D^*_s-D_s=144\MeV$
suggests the splitting between $cc$ states 
$\half(D^*_s-D_s)(s/c)=23\MeV$.
The $S=1$ diquark will then be 6 MeV higher than the hyperfine 
average and
then also the estimate for the $T_{cc}$ energy comes 6 MeV higher.
But $T_{cc}$ is anyway unbound by almost 100 MeV.

\subsection{The centre-of-mass motion}

In our phenomenological estimation we took the kinetic energy of the 
relative motion of the light diquark against the heavy diquark in  
$T_{bb}$ 
to be equal to the  kinetic energy of the light diquark against the $b$
quark in $\Lambda_b$.  To evaluate the correction to the
kinetic energy, we repeated the calculations carried out in \cite{BS1} 
where 
the $BD$ parameter set was used.

For the orbital part of the $T_{bb}$ wave function we took a simple 
ansatz 
where the wave functions of heavy diquark, light diquark and their
relative motion are Gaussians with widths $b_i$ which are chosen
so that the the energy of $T_{bb}$ is minimal
\begin{eqnarray}
R&=&\exp(-(r_1^2/2b_1^2+r_2^2/2b_2^2+r_3^2/2b_1^3)), \nonumber
\end{eqnarray}
$$
b_1 = 0.59\fm,\qquad  b_2= 0.23\fm,\qquad b_3= 0.60\fm.
$$
Here $r_1$, $r_2$ and $r_3$ are the distances between the heavy quarks,
between the light quarks and between the centres-of-masses of the two
clusters, respectively. This wave function is symmetric in permutation
of the two light quarks and in permutation of the two heavy quarks \
so that the spin-colour part of the function  for 
$T_{bb}$ in the state with total isospin, spin and parity $ISP = 10^+$
must have the form:
$$
S_{qq}C_{qq}S_{\bar{b}\bar{b}}C_{\bar{b}\bar{b}} = 
(0)_{qq}(\bar{3})_{qq}
(1)_{\bar{b}\bar{b}}(3)_{\bar{b}\bar{b}}
$$

Since we expect that this is the most important configuration we 
neglect 
the contribution of other configurations in the calculation of 
the correction to the kinetic energy.
We also expect that a more elaborate wave function of the type 
\cite{BS1} 
which gives $\sim 30 MeV$ smaller mass of  $T_{bb}$ would not change 
this correction by more than a few MeV.

The calculated kinetic energy of relative motion of 
$(\bar{b}\bar{b}) -- (qq)$
is 253 MeV. This kinetic energy should be smaller than the 
corresponding
kinetic energy of $b -- (qq)$ in $\Lambda_b$ used in our estimation
by a factor
\begin{eqnarray}
\frac{1/2q+1/b}{1/2q+1/2b}=1.06 .\nonumber
\end{eqnarray}
Therefore the mass of $T_{bb}$ should be for 15MeV
smaller than the phenomenologicaly estimated mass.

We also confirmed that the width of the  spatial wave function 
of the heavy diquark in $T_{bb}$ 
(calculated in the approximation with one Gaussian) is the same 
as in an isolated system of two $b$ quarks. This supports our
assumption that the presence of light quarks in dimeson does not
change  the  binding energy  of the heavy diquark (the change is less 
than
1 MeV).

\subsection{The finite size of the heavy diquark}

We again assume that the most important configuration in $T_{bb}$ 
is one
with antitriplet colour function for the light diquark and triplet 
colour 
function for heavy diquark. This makes $T_{bb}$  very similar to the 
baryon 
$\Lambda_b$ which we use in our phenomenological estimation. Here we 
calculate the correction to the heavy-light interaction energy because 
of the finite size of the heavy diquark. If we make the limit 
$b_2 \to 0$  
in the wave function defined in the previous subsection 
the binding energy is by 18 MeV less negative (for $BD$ parameters,
and a similar value for $AL1$). So this correction has 
the opposite sign as the correction due to the smaller kinetic energy 
and they almost cancel.

\subsection{The colour configuration mixing}

In $T_{bb}$ the light diquark and the heavy diquark can also be in 
colour 
(anti)sextet state. Such additional configurations are not present
in $\Lambda_b$ and can lower the energy of $T_{bb}$. This effect 
was calculated in Born-Oppenheimer approximation since such 
approximation
is very suitable for studing simultaneously the significance of sextet 
configurations as well as the tendency for  two-cluster configurations.
(next subsection).

The wave function of $T_{bb}$ is taken as a linear combination of
seven configurations $\Phi_i$ listed in table \ref{Configurations}
(for $\Phi_i$ we choose an orthogonalized basis)
$$
\Psi({\bf r}) =\sum_i  c_i({\bf r})  \Phi_i({\bf r}).
$$
We solved accurately seven coupled differential equations
\begin{equation}
-\frac{\hbar^2\nabla^2}{2m_{BB}}\,c_i({\bf r})
+\sum_j V_{ij}({\bf r}) c_j({\bf r}) = E c_i({\bf r})
\label{bornopp}
\end{equation}
where  $m_{BB}$ is the reduced mass of the two $B=\bar{b}q$ clusters
and the effective Born-Oppenheimer potential $V_{ij}({\bf r}) $ 
contains the kinetic energies of the light quarks as well as all 
interaction
energies. The calculation is done in the spirit of the generator 
coordinate
method (or resonating group method) except that $V_{ij}({\bf r}) $ 
is calculated with fixed heavy quarks.

With  the BD parameter set the binding energy is -70MeV if all seven 
configuration
from table \ref{Configurations} are used, and -45MeV if only 
the first three configurations (with colour triplet  wave functions)
are considered.
We find it plausible that colour sextet configurations, which were neglected in 
our phenomenological estimation reduce the mass
of  $T_{bb}$  by $\sim25\MeV$ also in an accurate calculation. The effect
is somewhat larger in the case of $T_{cc}$ due to the smaller mass
in the denominator of the spin-spin interaction which mixes the 
configurations, but the energy is still far above the threshold.

\begin{table}[htb]
\caption{%
Seven configurations for $T_{bb}$ with $ISP = 10^+$.
The spatial wave function of $T_{bb}$ is constructed from two delta 
functions 
for the two heavy quarks and two Gaussians for the two light quarks. 
The lower index  $\pm$ characterises the permutational symmetry 
of two identical particles.}
$$\left.
\begin{array}{ll}
M(b) \\
N(b)
\end{array} \right\} = \delta({\bf r}_b \pm {\bf r}/2 ),\qquad
\left.
\begin{array}{ll}
m(b) \\
n(b)
\end{array} \right\} = ( \frac{1}{\sqrt{\pi}k})^{-\frac{3}{2}}
 \exp\left(-\frac {({\bf r}_q \pm {\bf r} /2)^2}{2k^2}\right) .
$$
\begin{eqnarray*}
\phi_{1+} &=&\sqrthalf\left[m\left(q_1\right)m\left(q_2\right)+
n\left(q_1\right)n\left(q_2\right)\right] \\
\phi_{2+} &=&\sqrthalf\left[m\left(q_1\right)m\left(q_2\right)-
n\left(q_1\right)n\left(q_2\right)\right] \\
\phi_{3\pm} &=&\sqrthalf\left[m\left(q_1\right)n\left(q_2\right) \pm
n\left(q_1\right)m\left(q_2\right)\right] \\
\psi_{\pm} &=&\sqrthalf\left[M\left(b_1\right)N\left(b_2\right) \pm
N\left(b_1\right)M\left(b_2\right)\right] \\
\end{eqnarray*}
\label{Configurations}
\vskip\baselineskip
\centering
\begin{tabular}{l l l l l l l}\hline
 & & $qq$ &  & & $\bar{b}\bar{b}$ & \\
 &   $R_{q_1,q_2}$ &$S$ &$C$           &$R_{b_1,b_2}$ &$S$ &$C$   \\
\hline
$1$  &     $ \phi_{3+}$  &$0$ &$\bar{3}$    &$\psi_+$   &$1$ &$3$ \\
$2$  &     $ \phi_{1+}$  &$0$ &$\bar{3}$    &$\psi_+$   &$1$ &$3$ \\ 
$3$  &     $ \phi_{3-}$  &$1$ &$\bar{3}$    &$\psi_-$   &$0$ &$3$ \\
$4$  &     $ \phi_{3+}$  &$1$ &$6$    &$\psi_+$   &$0$ &$\bar{6}$ \\
$5$  &     $ \phi_{1+}$  &$1$ &$6$    &$\psi_+$   &$0$ &$\bar{6}$ \\ 
$6$  &     $ \phi_{2+}$  &$1$ &$6$    &$\psi_-$   &$1$ &$\bar{6}$ \\  
$7$  &     $ \phi_{3-}$  &$0$ &$6$    &$\psi_-$   &$1$ &$\bar{6}$ \\
\hline 
\end{tabular}\\
\end{table}

\subsection{Search for a two-cluster configuration 
(``molecule'' $BB^*$)}

At short distance, the colour triplet configurations 
$\Phi_{3+}, \Phi_{1+}$ and $\Phi_{3-}$ give a Coulomb-like attraction
between the two $B=\bar{b}q$  clusters while the colour sextet 
configurations give repulsion. At intermediate distances, on the other 
hand,
one can gain energy with a strong mixing between triplet and sextet
configurations. It turns out, however, that in the Born-Oppenheimer 
wave function of the previous subsection,
the amplitudes of colour sextet configurations are more than ten times 
smaller than the amplitudes of colour triplet configurations. This has 
the
consequence that the maximum probability of the relative motion
occurs at the origin.
Detailed calculations \cite{Janc} gave in fact no bound states
of $T_{bb}$ with a two-cluster ("molecular" or "covalent") structure.

\begin{figure}[htb]
\centering
\epsfig{file=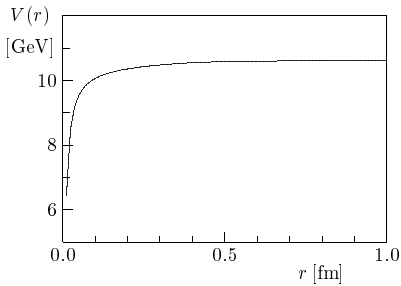,width=150pt}
\caption{%
The adiabatic Born-Oppenheimer potential for the $BB^*$ system}
\label{BO}
\end{figure}

In Fig. \ref{BO} we show the "adiabatic' Born-Oppenheimer
potential, where at each distance ${\bf r}$ separately the matrix
$V_{ij}$ has been diagonalized. This is not an accurate approximation,
but it demonstrates well the point that no second minimum appears 
in the $BB^*$ effective potential. Of course, the actual accurate
calculations of eigenenergies were performed with the full set of 
equations (\ref{bornopp}).

\section{Conclusion}

Assuming the equality of the wave function
of two light quarks in the dimeson and in $\Lambda_b$ as well as
the similarity between the subsystem of two heavy quarks in 
the $BB^*$ dimeson and  the quark-antiquark system in botomium 
(with scaled interaction strength)  we predict
a binding of  the $T_{bb}=BB^*$ dimeson of about -100 MeV. 
Analogous comparison with $J/\psi$ and $\Lambda_c$ predicts the dimeson
$T_{cc}=DD^*$ to be unbound.

The binding  energy of the $BB^*$ dimeson of about -100 MeV and no
binding for $DD^*$ agrees well
with some previous calculations \cite{SB3,BS1} in the consituent quark
model, since their model parameters reproduce well $\Upsilon$,
$J/\psi$, $\Lambda_b$ and $\Lambda_c$. On the other hand, model 
calculations which used unrealistically strong pion-exchange interacton
yielded very different predictions for the dimesons (many bound states);
they were of course irrelevant since they would strongly overbind 
$\Lambda_b$ and $\Lambda_c$ too.

Experimental verification of our predictions would be a challenge
since it would support the ``$V_{QQ}=\half V_{Q\bar{Q}}$ rule''

\appendix
\section{Input data and model parameters}
In order to be unambiguous we list in table \ref{Masses}
the baryon and meson masses
which we use as input for our phenomenological estimate.

\begin{table}[htb]
\caption{%
Masses of some baryons and mesons in MeV. 
Third and sixth columns give the hyperfine average of the vector meson
and the pseudoscalar meson above it}
\label{Masses}
\vskip\baselineskip
\centering
\begin{tabular}{l l l l l l}\hline
$\Lambda_b$     &  5624    & & $\Lambda_c$	&  2284.9  &	\\
                &          & & $\eta_c   $	&  2979.8  & 	\\
$\Upsilon $	&  9460.4  & & $J/\psi   $	&  3096.9  &   
3067.6   \\
$B_s      $    	&  5369    & & $D_s      $      &  1969    &    \\
$B_s^*    $	&  5416    & 5404\phantom{0000} &$D_s^* $ &  2112    
&  2076 \\
$B        $    	&  5279    & & $D        $      &  1867    &    \\
$B^*      $  	&  5325	   & 5314  &     $D^*      $ &  2008    
& 1973  \\
$K        $     & \phantom{0}495.7 &\\
$K^*      $     & \phantom{0}893.9 & \phantom{0}794.3 \\
\hline
\end{tabular}\\
\end{table}           

We also list in tables \ref{Mass} and \ref{Choice} 
the model parameters of Bhaduri et al. \cite{SB1} and of 
Silvestre-Brac and
Semay \cite{SB2} which we need in Sect.3 for our refinements. 
To unify the
notation we write down the form of the effective quark-quark 
interaction

\begin{eqnarray*}
V &=& -\frac{\vec{\lambda_i}}{2}\cdot\frac{\vec{\lambda_j}}{2}  
\left\{- \frac{\alpha_s}{r} + \kappa + \lambda r +
\frac{\tilde{\alpha}_s\,2\pi(\hbar c)^3}{3\,m_im_j c^4}\,
{\bf\sigma}_i\cdot{\bf\sigma}_j \,\delta^3(\vec{r})   \right\}\\
\delta^3(\vec{r}) &\to& \frac{\exp(-r/r_0)}{4\pi r r_0^2}\quad
{\rm or}\quad \frac{\exp(-r^2/r_0^{\prime 2})}{(\sqrt{\pi}r_0')^3} ,
\qquad r_0'=A\left(\frac{2m_im_j}{m_i+m_j}\right)^{-\nu} .
\end{eqnarray*}

\begin{table}[htb]
\caption{%
Choice of quark masses}
\label{Mass}
\vskip\baselineskip
\centering
\begin{tabular}{l l r r}\hline
Author  & & Bhaduri & Silvestre- \\
        & &         & -Brac      \\ \hline
$m_{u,d}\equiv q$         & MeV    & 337       & 315    \\
$m_s \equiv s$             & MeV    & 600       & 577    \\
$m_c \equiv c$             & MeV    & 1870      & 1836   \\
$m_b \equiv b$             & MeV    & 5259      & 5227   \\
\hline
\end{tabular}\\
\end{table}           

\begin{table}[htb]
\caption{%
Choice of parameters for the quark-quark interaction}
\label{Choice}
\vskip\baselineskip
\centering
\begin{tabular}{l l r r}\hline
Author  & & Bhaduri & Silvestre- \\
        & & $(BD)$  & -Brac $(AL1)$   \\ \hline
$\alpha_s$        & MeV fm &  77.0     &  75.0  \\
$\kappa$          & MeV    &-685.1     &-624.1  \\
$\lambda$         & MeV/fm & 705.7     & 628.4  \\
$\tilde{\alpha}_s$& MeV fm & 462.0     & 257,37 \\
$r_0$               & fm     & 0.4545    &        \\
$A$           & GeV$^{\nu-1}$&           &1.6553  \\
$\nu$               &        &           & 0.2204 \\
\hline
\end{tabular}\\
\end{table}           

\end{document}